\documentclass[allclo,superscriptaddress,eqsecnum,amsfonts,showpacs]{revtex4}
\usepackage{epsfig}

\newcommand{\be}{\begin{equation}}
\newcommand{\ee}{\end{equation}}
\newcommand{\bea}{\begin{eqnarray}}
\newcommand{\eea}{\end{eqnarray}}

\newcommand{\ep}{i\varepsilon}
\newcommand{\nn}{\nonumber}

\begin{document}

\preprint{ \parbox{1.5in}{\leftline{hep-th/??????}}}

\title{On the quark propagator singularity}

\author{V.~\v{S}auli}
\affiliation{CFTP and Departamento de F\'{\i}sica,
Instituto Superior T\'ecnico, Av. Rovisco Pais, 1049-001 Lisbon,
Portugal }

\begin{abstract}
Using the method of Fukuda and Kugo \cite{FUKKUG} the  continuation of Euclidean solution is performed to the timelike axis of fourmomenta. It is shown that 
assumed presence of the real simple pole in quark propagator is not in agreement with the solution. The simple pole disappears because of  the discontinuity in the 
resulting quark  mass function.
\end{abstract}

\maketitle
%

\twocolumngrid

Theoretical analysis of the singularities and the behaviour of  QCD Greens functions in the timelike momentum regime is still not well understood. Natural nonperturbative framework for the study of infrared properties of QCD Green's functions  is the formalism of Schwinger-Dyson equations (SDEs). For a review of recent progress in the whole QCD see \cite{FISCHER2006}.
Few possibilities of the quark propagator behaviour have been discussed in the literature.
As quarks are confined objects in nature one of the most natural expectation is the absence of a real pole in quark propagator. Some of the  models suggest that the real pole can be split into the complex ones. The complex conjugated poles have been phenomenologically appreciated in various studies. Furthermore, it has been recently realized in \cite{SAULI} that the structure of light  quark propagator is not reliably described by the Lehmann representation. The numerical analyzes exhibit  inefficiency of spectral representation  for correct description of  QCD chiral symmetry breaking. It is also known that when the strong interaction happen to  lead to the highly infrared singular interaction then the real quark  pole  naturally disappear \cite{MUN}.
  
However, the absence of pole in the quark propagator is not the only possible mechanism of quarks confinement and the existence/absence  of the real pole is still undetrmined \cite{ALSME2001}.  
The primary objective of this paper is to check the presence/absence of a real quark propagator pole  by using a simple generalization of  the  Fukuda-Kugo  timelike continuation  of ladder fermion SDE 
to the case of QCD, e.g. the asymptotic freedom is correctly taken into account through the known behaviour of running coupling. Using  usual assumptions we describe  contradicting results we obtained: The quark propagator does not exhibit the simple pole behaviour, instead, the mass function is largely discontinuous at observed real branch point.

The following conventions are used:  the  positive variables $x,y$  represent
 the square of momenta such that   $x=p^2$ for timelike 
momenta when $p^2>0$, while $x=-p^2$ for $p^2$ in spacelike region. Our metric is 
$g_{\mu\nu}=diag(1,-1,-1,-1)$. For purpose of  clarity  the mass function $M$ is labeled as 
$M_s$ in the spacelike region of fourmomenta and as $M_t$ when  evaluated for timelike
 fourmomenta (i.e., $M(p^2)=M_s(x)\theta(-p^2)+M_t(x)\theta(p^2)$).

The quark propagator $S$ can be conventionally  characterized by two independent scalars, the mass function $M$ and renormalization wave function $Z$ such that
\be
S(p)=\frac{Z(p)}{\not p-M(p)+\ep}\, ,
\ee
 noting the bare fermion propagator is $S_0=[\not p-m_0]^{-1}$, where
SDE for the inverse of $S$ reads
\bea \label{zdar}
S(p)^{-1}&=&\not p A(p^2)-B(p^2)
=\not p -m_0-
\nn \\
&&i g^2\int\frac{d^4q}{(2\pi)^4}\Gamma_{\alpha}(q,p)
G^{\alpha\beta}(p-q)S(q)\gamma_{\beta} \,
\eea
where simply $M=B/A$, $A=1/Z$ and $\Gamma$ is the full quark gluon vertex, $G^{\alpha\beta}$ represents gluon propagator, both  Greens functions satisfy their own SDE. 
To make the later  continuation  to the timelike regime more easily tractable 
we intruduce approximations that we believe do not change qualitatively feature of solution.
We will work in Landau gauge and take $A=1$. The importance of (un)presence of $A$  can be estimated from similar Euclidean studies. 
The so called Analytic Running Coupling (ARC) \cite{SOLMIL,SHISOL2007} is used 
to properly include the running of the QCD coupling. 
Noting that the effect of dynamical chiral symmetry breaking has been already studied  \cite{PAPNES} within ARC-SDE combined framework.

In  dressed one loop approximation   the following prescription of the SDE kernel is used
\be  \label{aprc}
g^2G^{\mu\nu}(k)\Gamma_{\nu}(q,p)\rightarrow 
4\pi\alpha(k^2,\Lambda)\frac{-g^{\mu\nu}+\frac{k^{\mu}k^{\nu}}{k^2}}{k^2+\ep}\, ,
\gamma_{\nu}  \, ,
\ee
with ARC   written via dispersion relation as 
\be  \label{rep}
\alpha(q^2,\Lambda_{QCD})=\int_0^{\infty} d\lambda\, \frac{\rho_g(\lambda,\Lambda_{QCD})}{q^2-\lambda+\ep}.
\ee 
 The correct one loop QCD running coupling at asymptotically large $-q^2$ is ensured 
when $\rho_g \simeq \rho_1$
\be \label{run}
\rho_1(\lambda,\Lambda_{QCD})=\frac{4\pi/\beta}{\pi^2-\ln^2{(\lambda/\Lambda^2_{QCD})}} \, ,
\ee
while the infrared  behaviour of ARC is modeled through the following modification of the spectral function:
\bea
\rho_g(\lambda)=(1+f_{IR}(\lambda))\rho_1(\lambda)
\nn \\
f_{IR}(\lambda)=Ce^{-(1-\frac{\lambda}{\Lambda^2_{QCD}})^2}.
\eea
where $\rho_1(\lambda)$ is given by (\ref{run}). Nonperturbative extra term $f_{IR}(\lambda)$
is chosen such that it does not affect ultraviolet asymptotics. Simultaneously it leads to the expected QCD scaling of infrared up and down quark masses $M_{u,d}(0)\simeq \Lambda_{QCD}$, where $C=15$ numerically.
 In order to define the kernel entirely we fix  $ \Lambda_{QCD}= 200MeV$
and $4\pi/\beta=1.396$, noting the later value corresponds with one loop ARC $\alpha(0)$ calculated  for three active quarks.

After making the trace,    SDE (\ref{zdar}) for the mass function reads
\bea
B(p^2)&=&m_0+3iC_A \int \frac{d^4q}{(2\pi)^4}\frac{4\pi\alpha((p-q)^2,\Lambda_{QCD})}{(p-q)^2+\ep}
\nn \\
&&\frac{B(q^2)}{q^2(q^2)-B^2(q^2)+\ep}\, ,
\eea
where $C_A=T_a T_a=4/3$ for $SU(3)$ group generators.

Using the integral representation (\ref{rep}) and  the following simple algebra 
\be
\frac{\alpha(z)}{z}=
\int d\lambda \frac{\rho_g(\lambda)}{\lambda}\left[
-\frac{1}{z}+\frac{1}{z-\lambda}\right]\, ,
\ee
we get after the Wick rotation and  angular integration  the resulting equation for $B$:
\begin{eqnarray} \label{reg}
&&B_s(x)=m_0+\frac{1}{\pi}
\int_0^{\infty}d y \, \frac{B_s(y)}{y+B_s^2(y)}
K(x,y)\,\, ,
 \\
&&K(x,y)=-\int_0^\infty d \lambda \frac{\rho_g(\lambda)}{\lambda}
 \left[ K(x,y,0)-K(x,y,\lambda)\right]\, \, ,
\nn \\
&&K(x,y,z)=\frac{2y}{x+y+z+\sqrt{(x+y+z)^2-4x y}} \label{reg2}\,.
\nn
\end{eqnarray}

\begin{figure}
\centerline{\epsfig{figure=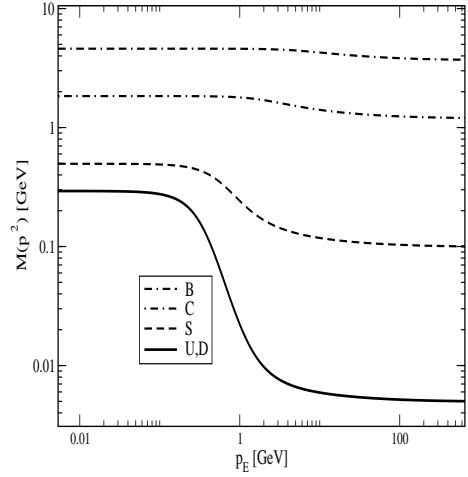,width=8truecm,height=7truecm,angle=270}}
\caption[caption]{Mass functions of the quarks. The thin lines are the results obtained by the integration through the absorptive part of $M$ in assumed dispersion relation as described in the text.} \label{spacequarks}
\end{figure}

The gap Eq. (\ref{reg}) contains potential UV divergence and can be renormalized after a suitable regularization. Since low relevance for our topic, the explanation is simplified and the renormalization issues are not discussed. In presented work we do not renormalize at all and solve the regularized gap equation. For this purpose  we follow hard cutoff scheme and introduce the upper integral regulator $\Lambda_H $, the numerical data has been obtained for $  \Lambda_H=e^8\Lambda_{QCD}$. Let us mention that we have explicitly checked that renormalization has only marginal numerical effect when comparing to the properly regularized solution solely. Within presented model  the mass function of the light and and heavy quarks has been calculated. The current bare masses were chosen to be
$m_{u,d}=5 \,MeV,\,  m_s=100\, MeV,\,  m_c =1\, GeV, m_b=3.5 \,GeV$. The resulting mass functions are plotted in  Fig. \ref{spacequarks}.

To make a continuation of    Euclidean Greens function to the physical timelike momenta is a rather cumbersome task. The assumptions are indispensable and not rarely if checked consequently, they  appear not justified in many physically important cases.  In what follows we will show that the assumption of Fukuda-Kugo continuation
in QCD is another case.

Changing trivially  the order of the integrations  we can immediately follow the Fukuda-Kugo prescription and write down the result for the timelike momentum. Continued SDE for the positive  timelike square of fourmomenta can be written in the following way:

\onecolumngrid
-------------------------------------------------------------------------------
\bea  \label{time}
&&B_t(x)=\hat{B}(x)-I(x)+ i\int_0^\infty  d\lambda \frac{\rho_g(\lambda)}{\lambda} 
\left[X(x;m^2,0)\Theta (x-m^2)-
X(x;m^2,\lambda)\Theta (x-(\lambda^{1/2}+m)^2)\right]
 \\
&& \hat{B}(x)=m_{0}+
\int_0^{\infty}d y \, \frac{B_s(y)}{y+B_s^2(y)}
K(-x,y)\,\nn \\
&&I(x)=\int_0^\infty \frac{d\lambda}{\pi} \frac{\rho_g(\lambda)}{\lambda}
P.\left[\int_0^{x}d y \, 
X(x;y,0) \Theta(x)
-  \int_0^{(\sqrt{x}-\lambda)^2}d y \, 
X(x;y,\lambda) \Theta(x-\lambda)\right]\frac{B_t(y)}{y-B^2_t(y)+\ep}
\nn \\
&&K(-x,y)=\int \frac {d\lambda}{\pi} \frac{\rho_g(\lambda)}{\lambda}
K(-x,y,0)-K(-x,y,\lambda)
\nn \\
&&=\int_0^\infty \frac {d\lambda}{\pi}\rho_g(\lambda)
 \frac{-\lambda-\sqrt{(-x+y)^2+4x y}+\sqrt{(-x+y+\lambda)^2+4x y}}
{-2x\lambda}
\, ,
\eea
-----------------------------------------------------------------------------------
\twocolumngrid
where the  function $K(a,b,c)$ is defined in Eq. (\ref{reg2}),
$m$ is a pole mass,
and the function $X$ is defined as
\be \label{vidle}
X(x;y,z)=\frac{\sqrt{(x-y-z)^2-4 yz}}{x}\, .
\ee
  
Eq. (\ref{time}) represents the integral equation for the mass function $ B_t(x)$ defined at the timelike axis $x$. It consists of the dominant  term $\hat{B}(x)$  represented by regular integral which including the mass spacelike regime defined function $B_s$.  It also contains the principal value integral that includes the complex function $B_t$  itself. As usually, absorptive part of $B_t$ is generated when crossing the branch point.

\begin{figure}
\centerline{\epsfig{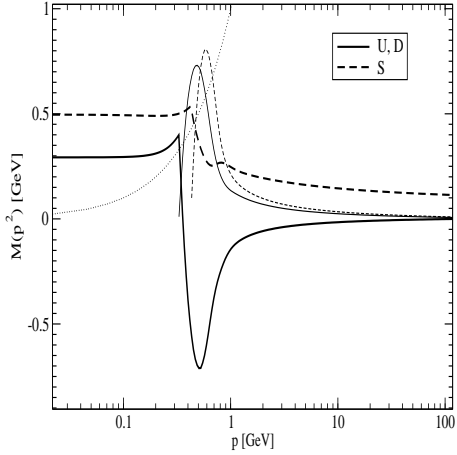}}
\vspace{1cm}\caption[caption]{ U,D,S quark mass functions for positive $p^2$. Thick (thin) lines represent the real (imaginary) parts. The dotted line is liner function in $p$.  } \label{timelightquark}
\end{figure}
The gap equation has been solved by the standard numerical  iteration. To check a numerical stability several integrators
 have been used to perform the integration numerically, e.g. the Gaussian and Simpson ones.   
The observed form of the 
discontinuity in numerically obtained quark mass function  strongly indicates that the singularity of quark propagator is softened such that
\be \label{lim}
lim_{p^2\rightarrow m^2} (p^2-m^2)S(p)=0\, ,
\ee 
which is in  contradiction with the assumption. The limit (\ref{lim}) is  well defined since the left and right limits  coincide. Note taht this is not  a case of $S$ because of discontinuity in $M_t$. The function $M_t$ becomes complex above the $m$ with $Im M$ starting from zero at this point. However, the right limit, if considered for the real part of $M$, is a bit model dependent, however we should stress that it has never leaded to the singular  propagator. There are two possible realizations observed numerically which spectacularly  differ  by the behaviour of $M_t$ in the right vicinity of $m$. Using a large UV cutoff one can find that the function $M_t$  cuts the linear function $p$  for some $p>m$ but  $Im M_t$ is always nonzero at that point. 
The second possibility that the inequaility $p-M_t(p)\not= 0$ is hold for any $p$ is observed for low scale cutoff. Such situation is displaed in Fig.4 for the case $\Lambda=5GeV$  and light quark
mass $m=5MeV$. Of course, the later case is  not real QCD, but this is rather qualitative test of cutoff independence. As mentioned the left limit of $M_t$ is less affected by the model changes, here it has been lowered about 50MeV. Thus in any case, one can conclude that the shape and perhaps complexity of the function $M_t$ in the vicinity of its discontinuity prohibit the  appearance of the real pole in the dressed quark propagator.

Termed in other way, we obtain finite quark propagator    
with discontinuous mass function at the point where the ordinary threshold would be expected.
If one need, the discontinuity can be classified
\be
lim_{p^2\rightarrow m_-^2} Tr S(p)\simeq \frac{-1}{\Delta},
\ee      
where $\Delta=B_t(m_-)-m$ is the "left discontinuity" of the function $B_t$ at the  point $m$.
The point $m$ is just defined by this discontinuity. Its numerical value   has been searched by
minimization of $\Delta$ during the iteration procedure for each quark flavor separately. The  numerical values we have found start from approximately $\Delta_{u,d}\simeq 100 MeV$ for vanishing current masses and slowly grows up to be  $\Delta_{c,b}\simeq 400 MeV$ for heavy quarks. The ratio $\Delta/m$ decreases for heavier flavors, which is agreement with expected supression of  nonperturbative effects in the case of heavy quarks.
\begin{figure}
{\epsfig{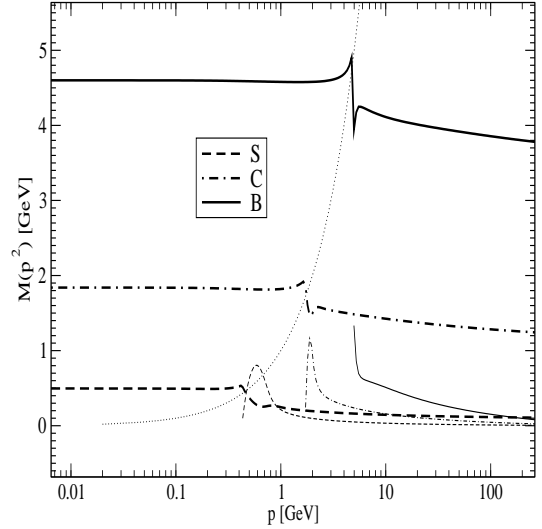}}
\vspace{1cm}
\caption[caption]{ C,B and S quark mass functions for positive $p^2$. Thick (thin) lines represent the real (imaginary) parts. The dotted line is liner function in $p$.   } \label{heavytimequark}
\end{figure}
The numerical solutions are displayed in Fig. \ref{timelightquark} for the light quarks $u,d,s$ and 
in Fig. \ref{heavytimequark} for the heavy flavors, in the later case, the mass function for the strange quark is added for comparison. 

\begin{figure}
\centerline{\epsfig{figure=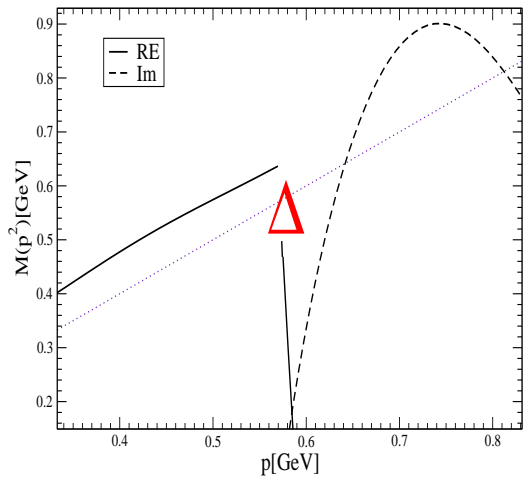,width=8truecm,height=8truecm,angle=270}}
\caption[caption]{ The detailed picture of discontinuity, the details are described in the text.} \label{details}
\end{figure}

In a strict sense the observed solutions should not  be taken seriously, since based on  wrong assumptions. On the other hand, the timelike continuation up to the discontinuity point $m$ should  be trustworthed. Then we can conlude that there is strong indication of disappearance of pure real pole at all, no matter what the continuation above the point $m$ is. The observation of mass gap $\Delta$ is in agreement with the confinement of quarks. Hadrons can never dissociate into the free quarks. A development of a new techniques for  more complete study of timelike behaviour of QCD Greens function are beeing looked for.

\onecolumngrid


\end{document}